# Commissioning of the CMS High Level Trigger

L Agostino[a], G Bauer[b], B Beccati[c], U Behrens[d], J Berryhil[e], K Biery[e], T Bose[f], A Brett[e], J Branson[g], E Cano[c], H Cheung[e], M Ciganek[c], S Cittolin[c], J A Coarasa[c], B Dahmes[h,i], C Deldicque[c], E Dusinberre[g], S Erhan[k], D Gigi[c], F Glege[c], R Gomez-Reino[c], J Gutleber[c], D Hatton[d], J Laurens[c], C Loizides[b], F Ma[b], F Meijers[c], E Meschi[c], A Meyer[d], R K Mommsen[e], R Moser[c,l], V O'Dell[e], A Oh[c,m], L Orsini[c], V Patras[c], C Paus[b], A Petrucci[g], M Pieri[g], A Racz[c], H Sakulin[c], M Sani[g], P Schieferdecker[c,n], C Schwick[c], J F S Margaleff[c], D Shpakov[e], S Simon[g], K Sumorok[b], A S Yoon[b], P Wittich[a], and M Zanetti[c,1]

[a]*Cornell University, Ithaca, NY, USA*
[b]*Massachusetts Institute of Technology, Cambridge, Massachusetts, USA*
[c]*CERN, Geneva, Switzerland*
[d]*DESY, Hamburg, Germany*
[e]*FNAL, Chicago, Illinois, USA*
[f]*Boston University, Boston, USA*
[g]*University of California, San Diego, San Diego, California, USA*
[h]*Lawrence Livermore National Laboratory, Livermore, California, USA*
[k]*University of California, Los Angeles, Los Angeles, California, USA*
[i]*University of Minnesota, Minneapolis, Minnesota, USA*
[l]*University of Technical University of Vienna, Vienna, Austria*
[m]*University of Manchester, England*
[n]*Universitaet Karlsruhe, Germany*

E-mail: `marco.zanetti@cern.ch`

**Abstract.** The CMS experiment will collect data from the proton-proton collisions delivered by the Large Hadron Collider (LHC) at a centre-of-mass energy up to 14 TeV. The CMS trigger system is designed to cope with unprecedented luminosities and LHC bunch-crossing rates up to 40 MHz. The unique CMS trigger architecture only employs two trigger levels. The Level-1 trigger is implemented using custom electronics, while the High Level Trigger (HLT) is based on software algorithms running on a large cluster of commercial processors, the Event Filter Farm. We present the major functionalities of the CMS High Level Trigger system as of the starting of LHC beams operations in September 2008. The validation of the HLT system in the online environment with Monte Carlo simulated data and its commissioning during cosmic rays data taking campaigns are discussed in detail. We conclude with the description of the HLT operations with the first circulating LHC beams before the incident occurred the 19$^{th}$ September 2008.

---

[1] Corresponding author.

## 1. Introduction

The CMS detector [1, 2] is now built and in its final commissioning phase, preparing to collect data from the proton-proton collisions to be delivered by the Large Hadron Collider (LHC), at a centre-of-mass energy of up to 14 TeV. The CMS experiment employs a general-purpose detector with nearly complete solid-angle coverage, which can efficiently and precisely measure electrons, photons, muons, jets (including tau- and b-jets) and missing transverse energy over a wide range of particle energies and event topologies. These characteristics ensure the capability of CMS to cover a broad programme of precise measurements of Standard Model physics and discoveries of new physics phenomena.

The trigger and data acquisition system must ensure high data recording efficiency for a wide variety of physics objects and event topologies, while applying online very selective requirements.

The CMS trigger and data acquisition system [3, 4] is designed to cope with unprecedented luminosities and interaction rates. At the LHC design luminosity of $10^{34} cm^{-2} s^{-1}$, and bunch-crossing rates of 40 MHz, an average of about 20 interactions will take place at each bunch crossing. The trigger system must reduce the bunch-crossing rate to a final output rate of O(100) Hz, consistent with an archival storage capability of O(100) MB/s.

Only two trigger levels are employed in CMS. The first one, the Level-1 Trigger (L1T) [3], is implemented using custom electronics and is designed to reduce the event rate to 100 kHz. The second trigger level, the High Level Trigger (HLT), provides further rate reduction by analyzing full-granularity detector data, using software reconstruction and filtering algorithms running on a large computing cluster consisting of commercial processors, the Event Filter Farm.

In this paper we discuss the features of the CMS High Level Trigger and its implementation in the Event Filter System. We then report details about the HLT validation and commissioning, both with simulated LHC data and with cosmic ray data. We conclude by describing the operations of the HLT with the first LHC beams in September 2008.

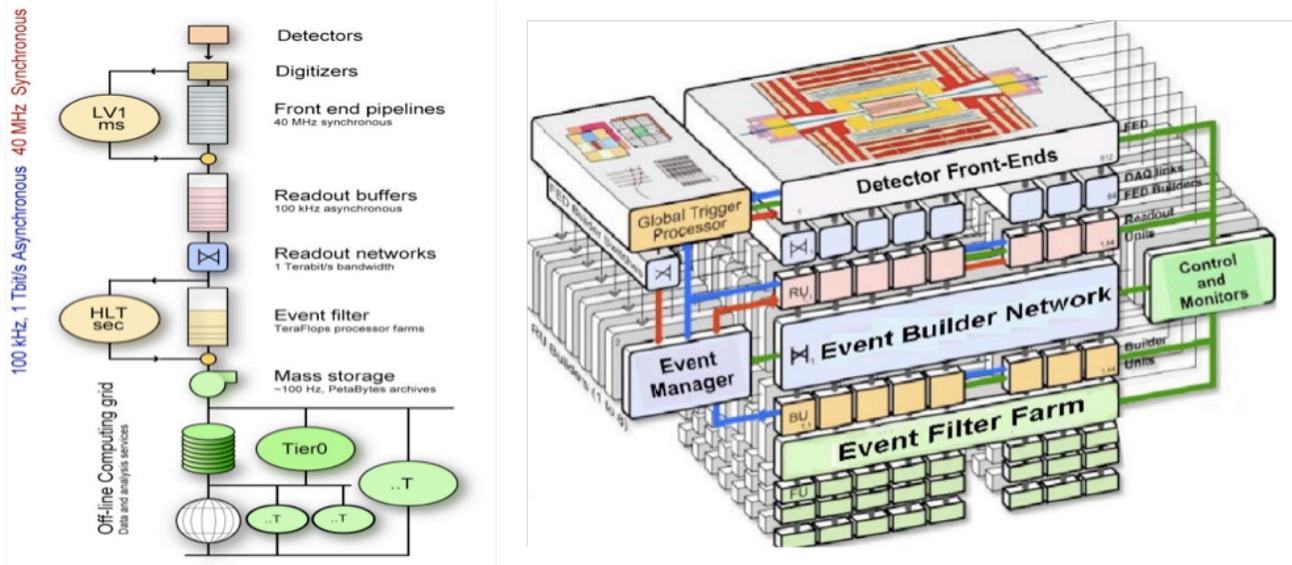

**Figure 1.** Left: schematic view of the CMS Trigger and DAQ architecture. Right: representation of the DAQ system from the front ends to the storage system. The development in the third coordinate corresponds to the implementation of several independent DAQ slices.

## 2. Trigger and DAQ system layout

The layout of the CMS trigger and DAQ architecture is sketched in figure 1. Given the high event rate at the nominal LHC luminosity, only a limited portion of the detector information from the calorimeters and the muon chambers is used by the L1T system to perform the first event selection, while the full granularity data are stored in the detector front-end electronics modules, waiting for the L1T decision. The overall latency to deliver the trigger signal (L1A) is set by the depth of the front-end pipelines and corresponds to 128 bunch crossings. The L1T processing elements compute the physics candidates (muons, jets, e/γ, etc.) based on which the final decision is taken. The latter is the result of the logical OR of a list of bits (up to 128), each corresponding to a selection algorithm. All the trigger electronics boards are fully programmable as in particular is the set of selection algorithms, the L1T menu [5].

The full detector data (~1MB) corresponding to the events accepted by the L1T are read out by the DAQ system at a rate up to 100 kHz. The event building [6], i.e. the assembling of the event fragments coming from each detector front-end module, takes place in two stages. First, the front end data are assembled into larger fragments (super-fragments) which are then delivered to Readout Units (RU) in eight different and independent sets (DAQ slices) in a round-robin scheme, such that all super-fragments of an event are delivered to the same DAQ slice. In each DAQ slice the super-fragments are managed by the Event Builder through a complex of switched networks (Gigabit Ethernet) and passed to event buffers (Builder Units, BU) where they are finally assembled into complete events. From the BU, the events are handed to the Filter Units (FU), the applications which run the actual High Level Trigger reconstruction and selection. Events accepted by the HLT are forwarded to the Storage Managers (SM), two for each DAQ slice, which stream event data on disk and eventually transfer raw data files to the CMS Tier-0 computing center at CERN for permanent storage and offline processing. More details about the last stage of the data acquisition process will be given in the next section.

## 3. Event Filter System

In this section, the design features and the implementation details of the Event Filter System as of the time of the first LHC beams are described.

*3.1 Architecture*

The Event Filter Farm currently consists of 720 nodes (BUFU) where the BU and the FU applications are both executed. The internal architecture of the BUFU nodes is shown in figure 2 and described in detail in Ref [7]. A choice has been made to couple in the same node the application that buffers and formats the event data, the Builder Unit, and the actual execution of the HLT reconstruction and selection.

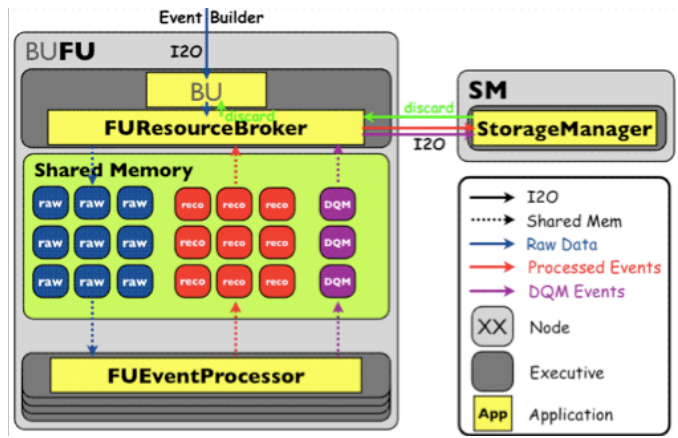

**Figure 2.** Internal architecture of a BUFU node. O(1000) of them comprises the Event Filter Farm

A dedicated application, the Resource Broker (RB) takes care of exchanging data with the DAQ (requiring high bandwidth I/O), decoupling the execution of the physics algorithms (CPU intensive), performed by the Event Processors (EP), from data flow. This allows each Filter Unit to continue operation, recover the content of the problematic events, and forward them to be stored unprocessed.

A key feature of this design is the execution in parallel of several independent filtering processes, which benefits from the multi-core architecture of the most recent CPUs. On demand by the RB, the event data are fed into each EP from a queue in a shared memory buffer (*raw cells*) and the processing happens asynchronously. The events accepted by the HLT selections are managed by the RB in a separated shared memory buffer (*reco cells*) and sent to the Storage Manager. Monitoring data are handled in complete analogy to event data. Data quality monitor (DQM) software modules are executed together with the HLT reconstruction and selection algorithms by the EP. The information produced (typically histograms) are shipped to the SM through additional shared memory buffers (*DQM cells*).

*3.2 Control*

The various components of the system are configured and controlled by the CMS Run Control and Monitoring System (RCMS [8]), providing a hierarchical control structure (a tree of Finite State Machines) for all the detector subsystems involved in the data acquisition, i.e. the DAQ system itself, the L1T, the DQM, and all the sub-detectors. By means of web-based applications, RCMS distributes commands, and monitors the state of all system components. In this way, a single web graphical user interface is used to operate the complete DAQ system of the experiment.

When Run Control initializes the DAQ system, each application process is started on its assigned computing node. All Filter Farm applications implement a common State Machine, which is defined by the three states Halted (initial state), Configured, Enabled, and transitions between these states. In addition, an Error state indicates an irrecoverable failure of the corresponding application. Run Control can be instructed to send control messages to each application and trigger state transitions, associated with a set of actions and an expected target state. The state of each application is monitored at all times. Controlled components actively notify modifications of their state, which are accounted for. Unexpected or missing notifications are handled appropriately.

*3.3 HLT Configuration*

The HLT reconstruction and selection is implemented in the same software framework used for the offline reconstruction and analysis [9]. This provides a high degree of flexibility in the use of sophisticated algorithms developed for offline reconstruction, while allowing further developments and modification of the said algorithms to be easily integrated in the online selection.

The framework uses a modular architecture, enforcing interfaces for the implementation of the different functionalities: input, reconstruction, filtering, access to non-event data, other services (such as logging, monitoring, etc.), and output. The modules to be executed at runtime are defined by means of a configuration document and loaded, instantiated, and configured using a plug-in mechanism at initialization time.

The configuration document provides the actual definition (instance) of each module specifying the values of all of its relevant parameters. Several instances of the same module (e.g. different definition of a tracking algorithm) can be present in the configuration. The modules' instances are grouped into "paths". In each path, the order of the instances corresponds to the order of execution.

The HLT configuration, in the following referred to as the HLT table or menu, is composed of a set of (trigger) paths, each consisting of (instances of) reconstruction and filtering modules. Each trigger path typically addresses a specific physics object selection (e.g. di-muon events). In analogy to the L1T, the acceptance for permanent storage of an event requires it to be selected by at least one of the HLT trigger paths. In order to guarantee reproducibility of the results, and the complete traceability of events accepted by the HLT, all paths are executed for all events. On the other hand, the execution of a path is interrupted if the event being processed does not fulfill the conditions required by a given filtering module.

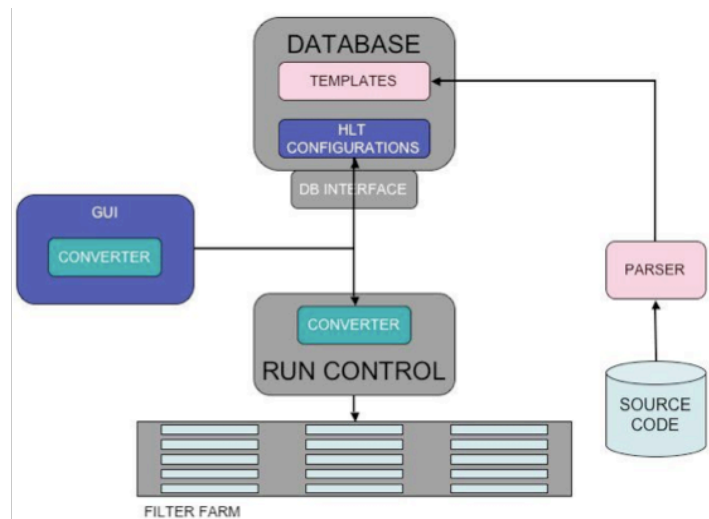

**Figure 3.** Schematic representation of HLT configuration workflow

The creation of an HLT configuration and its deployment to all Filter Farm nodes is based on a robust database design [10]. The definition of a configuration is abstracted in a relational database schema: for a given software release, a thin code parsing layer is employed to discover the relevant modules for the HLT execution and all their parameters, and store them as templates in the database. HLT configurations based on these modules' templates can be created and stored with a dedicated graphical user interface (figure 3).

Configurations can be retrieved and transformed into several useful representations, most notably the one used to configure the reconstruction framework, to be distributed to each filter node via the Run Control layer. Formatting and deploying of the configuration are decoupled from the database schema, allowing the target configuration grammar to evolve independently.

### 3.4 Calibration and Condition Data

Non-event information such as detector conditions, configuration and calibration are essential to properly interpret the detector measurements on which are the base of the online selection. The CMS reconstruction framework [11] retrieves these data from a relational database through an object-to-relational mapping tool (POOL-ORA[12]) relating C++ objects to the relational database schema. A web based service, FroNTier [13], is used to distribute over the network the information from the central database service to the various clients (reconstruction jobs) in HTTP format [14].

In this respect the Event Filter Farm poses stringent requirements: the conditions and calibration information, conservatively estimated to be 100 MB, have to be loaded simultaneously at the beginning of every run into O(10000) processes (Filter Units) running on ~1000 nodes in the shortest time possible, since any delay results directly in data taking dead time. To deliver O(1) TB of data to the clients, a hierarchy of Squid proxy/caching servers are deployed between each client and a central redundant FroNTier server connecting to the database. Specifically, a squid is deployed on every Filter Farm node providing a fast simultaneous start (few seconds) of the Event Processors in the case the conditions data are already cached and do not need to be updated. Six further squid tiers, with each squid fanning out four squids in the next tier, are used in the current configuration, allowing the loading of new conditions in less then a minute.

### 3.5 Data Storing and Transfer to Tier 0

The events accepted by the HLT selection are delivered to the Storage Manager applications, which take care both of writing the data on local disks and transferring them to the CMS Tier 0 at CERN.

In its final configuration, the storage system will consists of 16 nodes (two for each DAQ slice) connected to 8 SATABeast$^{TM}$ disk arrays with a maximum expected capacity of 320 TB. Data are transferred from the filter farm nodes to the SM nodes via a Gigabit Ethernet switch with a steadily sustainable throughput of the order of 2 GB/s. The connection to the Tier 0 exploits a 10 Gbit redundant optical link, for an overall bandwidth from the detector front-ends to the Tier 0 of about 800 MB/s. The same binary protocol used for event building is also used for data transactions between the Filter Units and the Storage Manager. Each transaction must be acknowledged by the SM to the Resource Broker before events can be discarded from the event builder, thus allowing an indirect back-pressure mechanism on the L1 trigger in case of large fluctuations of the accept rate.

Routing of individual event data to files by the SM is driven by the definition of output streams in the HLT configuration. Several streams, corresponding to distinct sets of data files, can be enabled in order to group together events according to their offline usage. Each data stream receives events from a predefined set of trigger paths. The same path can however feed several different streams (as explained later, this allows for instance the monitor of that HLT path).

Since the reconstruction software framework requires event data files (both online and offline) to contain homogeneous information, different HLT streams must also be defined for samples requiring special or reduced event data. As an example, the event content of the data samples to be used for the offline physics analysis (physics stream) consists usually of the complete collection of detector and trigger raw data, as well as of the L1 and HLT selection results. On the other hand, detector calibration can benefit from the highest possible collection rate. Hence, calibration and alignment streams only store a portion of the raw data or dedicated collections of reconstructed objects, allowing higher HLT accept rates than the physics stream.

Data transfer from the CMS experimental site to the Tier 0 computing centers and the regional Tiers is driven in the first instance by the definition of the streams and primary datasets. At the Tier 0 site, data streams received from the experimental site are prioritized according to their specific workflows: calibration and alignment data are processed first in order to compute the conditions to be used to best reconstruct the physics data stream. Analogously, events featuring physics properties of compelling interest, the "Express" stream, are processed as soon as the corresponding data files are received. Within a stream, sets of paths performing similar selections can be further grouped to define the so called "primary datasets", the data samples used to address the various offline physics analyses. Primary datasets are also produced at the Tier 0 by splitting the data streams based on the result of the HLT selection.

## 3.6 Error Handling

Despite all efforts to debug and validate the HLT code before it goes in production, pathological detector conditions, or genuine bugs in portions of the code that are only rarely executed, can result in a failure of the EP. The BUFU complex of applications is able to continue its operation undisturbed even in the event of failure of one or more EP processes. A failure can be detected directly as the consequence of a fatal execution error resulting in a transition to the error state of the application, or indirectly by the RB as a timeout of the corresponding assigned resource in case of unexpected runtime exceptions resulting in a crash. In the latter case, an error message is generated, and it is the responsibility of the control system to ensure that the corresponding process be properly cleaned up.

In both cases, provisions are made to dispose of the corresponding resource and discard it from the system. Detector raw data for the event that caused the error are handed over to the SM to be logged to an "error event" stream. The error stream is vital to collect real-life pathological events to be used as test samples for the validation of the HLT code.

Error messages generated by reconstruction code in the Event Processor are collected and analyzed using the error and log collection capabilities provided by the online framework (XDAQ [15]) and Run Control.

## 3.7 Data Quality Monitoring

Like the HLT event filter software, the online event data quality monitoring software is implemented in the standard offline software framework.

At CMS, real time data quality monitoring is done in two places, namely inside the HLT event filter applications as part of the HLT menu described above, and also downstream of the Storage Manager system. A sketch of the monitoring data flow is given in figure 4.

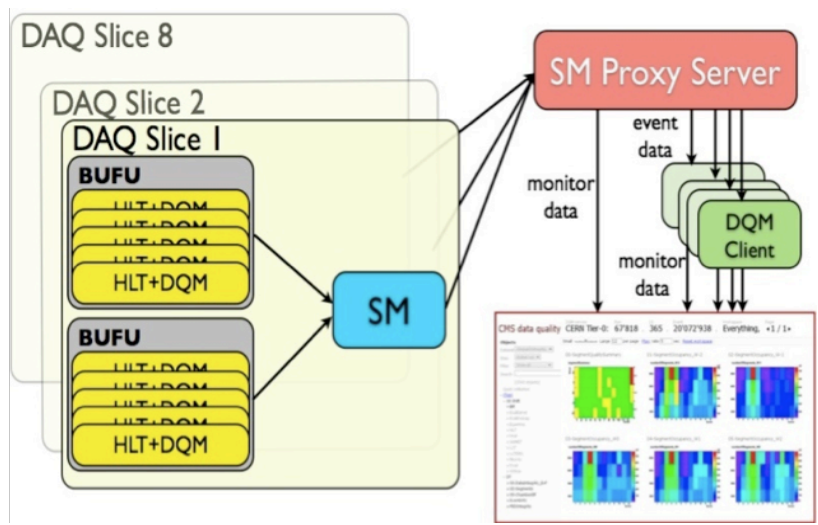

**Figure 4.** Schematic representation of the monitoring data flow

The monitoring information from the HLT is delivered to the SM through a queue of shared memory cells as described in section 3.1. The production of monitoring information inside the HLT event filter gives access to the full L1T accept rate (up to 100kHz), but is limited in terms of CPU power and memory, in order not to restrict the HLT resources required for event reconstruction and selection. In practice the HLT based DQM modules are restricted to fast checks of the correctness of the data formats (data integrity) and the monitoring of the L1 and HLT trigger system itself.

To compensate for the limitation in the complexity of the monitoring algorithms and the number of histograms at the HLT level, an additional system of dedicated monitoring applications has been implemented that processes events parasitically from a dedicated network event data web server downstream of the SM. These independent consumer applications receive and process events at typically 20 Hz and can be freely configured to perform very complex reconstruction and monitoring algorithms.

## 4. HLT Menu Development

Before describing in the next sections the validation and the commissioning of the High Level Trigger system, we recall in the following the motivations and the results of a detailed study [16] aimed to develop HLT menus with the expected physics and computing performance for LHC startup luminosities up to $O(10^{32})$ $cm^{-2}s^{-1}$.

The actual trigger performance will be measured and optimized with real data from collisions. Nonetheless our present best knowledge of the detector response and possible collider condition scenarios can be used to study and optimize trigger criteria, in view of adjusting them when real collisions and detector data will be available. The flexibility of the trigger system allows to introduce modifications in an efficient manner for optimal performance, adapted to the actual running conditions while taking data. The robustness of the algorithms which determine the trigger primitives also ensure that the system will not be too sensitive to small changes with respect to the expected conditions.

In preparation for data taking, the HLT menus are developed using fully simulated and reconstructed Monte Carlo events for the known Standard Model processes, dominated in rate by QCD events at low transverse momentum, also known as minimum bias events. The goal is to achieve a reduction factor of ~1000 on the input HLT rate, while keeping as high as possible efficiency for events of interest, with an average processing time per event of the single HLT instance not exceeding 50 ms [4]. To meet these requirements, the menus are structured as a set of trigger paths, each dedicated to select events with specific topologies and kinematics.

The CPU time required for the execution of the HLT algorithms in the filter farm is minimized by rejecting events as quickly as possible, using a limited amount of detector information. In each path, the filtering modules are embedded among the reconstruction sequences such that if the requirements are not matched, the rest of the path is not executed. In particular, every HLT path begins by requiring a specific L1T precondition (i.e. logical combination of a set of L1 bits). Moreover, only the parts of the detector pointed to by the L1T candidates (the physics objects candidates on which the L1T decision was based) are usually considered for further validation of the trigger object being scrutinized. Physical parameters (transverse momentum, pseudo-rapidity, azimutal angle) of the L1T candidates can also be used as starting point ("seeds") for the HLT reconstruction. The choice of considering only part of the detector information to reconstruct the event is obviously not mandatory but is driven by the need of saving processing time. The full detector data is anyway at disposal for use by the HLT algorithms if needed.

Repetition of identical operations is minimized. The framework guarantees that identical instances of the same reconstruction module present in several paths are executed only once. After each possible reconstruction step, a set of selection criteria, applied to the reconstructed objects, results in the rejection of a significant fraction of events, thus minimizing the CPU usage at the next step.

In the summer 2007, a complete HLT menu with these features has been produced as the candidate menu for physics runs at $L=10^{32}$ $cm^{-2}s^{-1}$. The plot in Figure 5 summarizes the performances of the candidate menu in terms of timing on a sample of minimum bias events passing the corresponding L1 trigger menu.

## 5. HLT online validation

In order to test and validate candidate HLT menus in realistic online conditions and estimate the required computing power, a playback system was put in place and used as a test bed.

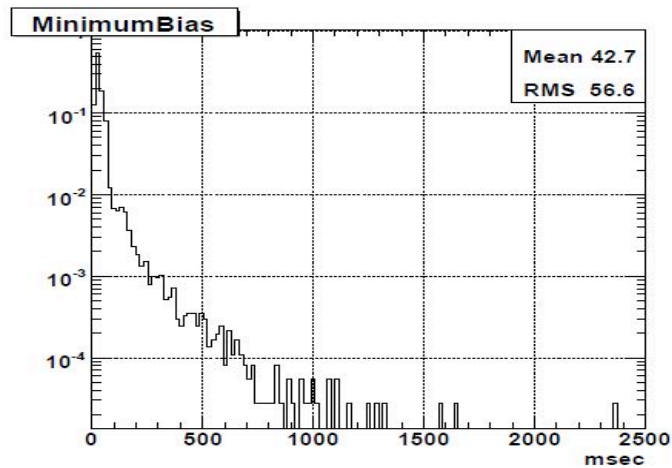

**Figure 5.** Timing profile for the HLT menu developed for LHC luminosity $L=10^{32}$ cm$^{-2}$s$^{-1}$ for a sample of QCD events passing the L1T selection. The results are obtained on a PC mounting 3 GHz Intel CPU [16]

The playback setup, shown schematically in figure 6, uses the same architecture of the BUFU nodes as in the full DAQ system, but in replacement of the events provided by the Event Builder Network (see figure 2), an application reads the events from a data file and passes them to an ad-hoc version of the BU (auto-BU). From this level on, the data exchange protocol is the same as described in Section 3.1. The auto-BU is configurable in such a way that once event data are read from the file, they can be re-played continuously.

Several BUFU nodes can be run in parallel, having one or more SM's receiving events from each of them. This allows to test the scalability of the system as well as to compare simultaneously the performances of different commercially available CPUs. For the validation of the HLT menu at $L=10^{32}$ cm$^{-2}$s$^{-1}$ [5] 20 BUFU nodes were employed. The frequency of the CPUs installed in those nodes ranged between 2 and 3 GHz (Intel® processors).

As input for the validation tests, a Monte Carlo sample of 20 million minimum bias events was produced and passed through the full detector simulation. The L1 menu decision was emulated in order to select only those events which would pass the L1 trigger, reducing the initial sample to 50,000 events. Realistic detector conditions were simulated, corresponding to the estimated uncertainties on the calibration constants as expected after collecting an integrated luminosity of 100 pb$^{-1}$. The mean processing time per event was 50 ms

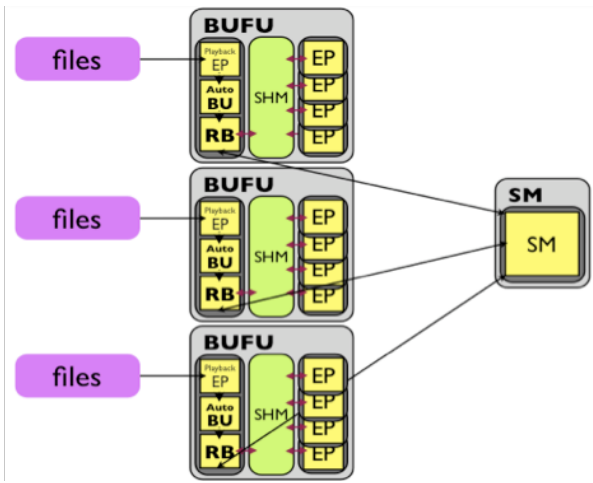

**Figure 6.** Layout of the playback setup of the Filter Farm used for the online validation of the HLT code and configurations. Events are readout from Monte Carlo event files by means of a dedicated application replacing the Event Builder Network. They are then passed to the Filter Units using the standard data exchange protocol.

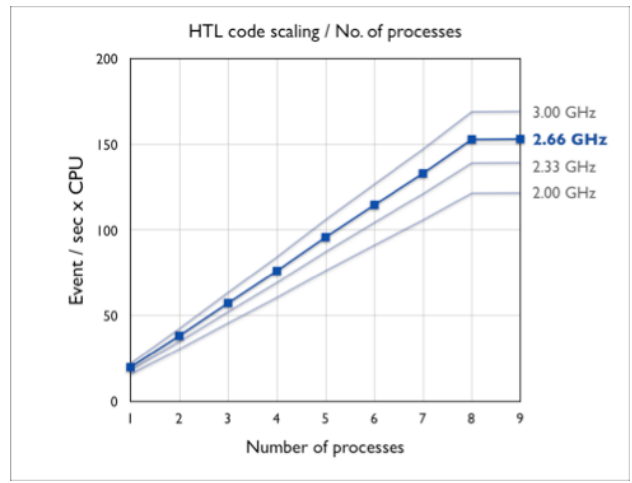

**Figure 7.** Event processed per second as a function of number of HLT instances running per node. Results are shown for CPU frequencies ranging from 2 to 3 GHz

matching well the initial design requirements. This measurement was obtained on a 2.66 GHz CPU.

The scaling properties of the HLT processing time as a function of the CPU frequency have been studied. As expected, the average number of events processed per second as a function of the CPU frequency increases linearly as shown in the plot of figure 7. The results have been obtained by comparing four Intel® processors. Recently, commercially available processors combine two or more independent cores into a single die, or more dies packaged together. The performance of multi-core processors running several instances of the HLT executable has also been tested. The plot of figure 7 shows the number of processed events per second as a function of the number of HLT instances running on a dual quad-core machine. The number of processed events increases almost linearly up to eight instances, when the processes occupy almost entirely the available CPU power[2]. When nine, or more, processes run in parallel the overall CPU power is equally shared among each process, yielding the same timing result achieved with eight processes.

The memory usage could be an issue in the case of multiple HLT instances running concurrently on the same node. The HLT menu under test was measured to demand not more that 600 MB of resident memory. A memory slot as large as 2 GB was assigned for each core.

The average power consumption is only mildly affected by multiple processes running in parallel on the same node. The increase of the power consumption when all the CPUs are fully used with respect to the idle state is 20%.

Given the results of the tests on the candidate HLT menu for $L=10^{32}$ cm$^{-2}$s$^{-1}$, it has been decided to base the Filter Farm on PC's mounting 2.66 GHz dual quad-core Intel® processors (*Clovertown*) and a total of 16 GB RAM. In October 2008 720 of these PC's were installed in the CMS counting room, commissioned and partly employed to run the HLT menus during CMS global data taking runs with cosmic muons (see Section VI).

Assuming a conservative factor of two in the average HLT processing time, corresponding to an average number of 10 events processed per second and per core, the CPU power currently available in the Filter Farm will be suitable for handling an HLT input rate as high as 60 kHz, beyond what is expected for the first year of LHC physics runs. An additional safety factor is provided by the fact that more recently developed HLT menus reduce the average processing time with respect to what was obtained previously. Moreover, the computing power of the CMS Filter Farm is expected to be doubled in size, with newer CPUs, before the start of LHC operations at high luminosity.

### 6. HLT commissioning with cosmic rays

Since the final assembly of the detector in the underground cavern in summer 2007, a constantly increasing fraction of the CMS sub-detectors has been commissioned and integrated with the trigger and DAQ systems. Global data taking campaigns recording cosmic muon events take place regularly with the goal to commission the experiment for the first beam data. In this context, the HLT system plays a pivotal role, providing selected data samples for commissioning of the different sub-detectors, and linking the online environment with the offline computing and analysis workflows. At the same time, the data taking exercises offer a unique opportunity to commission the functionalities and verify the performances of the High Level Trigger system. Among the various goals, the following items were addressed with special care:
- Sustainable input rate
- Stability, reliability and timing performances of the reconstruction and selection algorithms
- Interface with the L1 trigger
- Performance of the storage system
- Infrastructure for DQM
- Robustness against pathological event data and detector conditions

*6.1 HLT menus for cosmic ray data taking*

During commissioning with cosmic muons, ad-hoc code and configurations were employed in the HLT to facilitate the detector commissioning process. For example, at least one path in the HLT menu was configured to accept all physics events that fired the L1 trigger, while other HLT paths applied very simple filters (based for instance on the reconstructed muon direction or on the number of reconstructed tracks in the tracking system) and tag the events for easier offline processing and physics skimming.

---

[2] The other two processes running on the BUFU node, i.e. the (auto)BU and the RB require very little CPU power, less than 5% in total.

The rate of cosmic events in the CMS cavern does not exceed ~500 Hz. In order to stress the HLT system with a high input rate, Poisson distributed pseudo-random triggers have been generated and delivered together with the cosmic triggers by the L1T at frequencies as high as 60 kHz. Given the reduced capability of the Filter Farm in terms of computing power during the early detector commissioning phase (320 CPUs, 4 SMs), the random triggered events have not been fully reconstructed but just rejected on the basis of a high prescale factor. A small fraction of random trigger were accepted, tagged and streamed out separately from the cosmic muon events.

In order to commission the HLT reconstruction and selection code, menus of increasing complexity have been deployed online. In particular, a complete menu developed on Monte Carlo data for LHC collisions at $L=10^{30}$ cm$^{-2}$s$^{-1}$, has also been tested online during cosmic muon data taking. This menu, consisting of ~100 HLT paths, with more than 1000 different algorithm instances, proved to run steadily for several days. Due to the extremely simple topology of the events, the overall processing time per event was measured to be about 20 ms. Given the continuous changes in the data taking conditions (upgrades of the L1T menus, varying current in the superconducting solenoid, etc) in order not to interfere with the other ongoing tests, the complexity of the HLT menu was required to be limited. Nonetheless, in addition to the ad hoc trigger paths designed for specific detector studies, a substantial fraction of the trigger paths from the full menu were employed in most of the runs.

The first step of the HLT data processing is the unpacking of raw data. The algorithms performing this operation have to be robust against data corruption and missing information and at the same time monitor and log possible readout errors. In particular, the data from the Global Trigger [5], the electronic board performing the final L1T selection, need to be unpacked for every event for the L1T decision bits to be available for the HLT processing. During commissioning with cosmic rays, the (detector and L1T) data unpacking and path seeding mechanism (see section IV.C) have been extensively tested, requiring that every HLT path be seeded by one or more L1 bits. Both muon and calorimeter (jets and $e/\gamma$) L1 bits have been employed demonstrating the effectiveness of the interface between L1T and HLT. In particular, these tests have proven that L1 muon candidates provide seeds for high efficiency HLT muon reconstruction.

The event display [17] in Figure 8 shows a muon track reconstructed online by the HLT algorithms, seeded by the information generated by the L1 muon Trigger (Global Muon Trigger [18]). The flexibility granted by the employment of the same framework as the offline software eased the used in the HLT menus of dedicated reconstruction algorithms more suitable to cosmic muon events than the standard LHC reconstruction. The muon track in Figure 8 has for instance been fitted by a special tracking code developed to deal with particle trajectories not originating from the center of the detector (interaction point) and to reconstruct trajectories traversing the experiment from the top to the bottom as single tracks [19].

*6.2 HLT output, data streams and primary datasets*

The concurrent production of several data streams has been tested during the commissioning runs with cosmic rays. The physics events, i.e. those firing the L1T bits dedicated to trigger on cosmic muons crossing the detector, were grouped into a unique stream (named "*A*"). The information content of this stream consisted in detector raw data and trigger (L1T and HLT) results only. Once transferred to the CERN Tier-0, the data files corresponding to stream "*A*" were reformatted and split into several Primary Datasets accordingly to the HLT results. As an example, one of the most interesting datasets for the offline analysis was the one containing events with at least one track reconstructed in the silicon strip tracker by the HLT.

In addition to the physics stream, alignment and calibration streams, as foreseen for LHC collision data taking, have been produced. In particular one of these streams was composed by events generated by special trigger signals ("calibration" triggers) delivered at the time corresponding to LHC abort gaps once per orbit which each sub-detector uses in order to enable test-pulses sequences (e.g. laser beams on the calorimeter crystals to probe the evolution of their light yield). Another of these streams was composed of partial event data from L1 physics triggers further selected by the HLT in order to enhance events with specific properties.

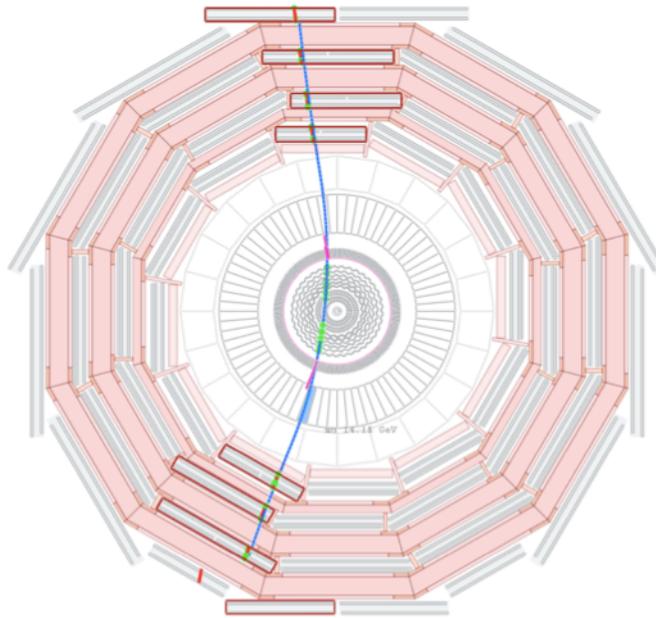

**Figure 8.** Event display of a cosmic muon reconstructed online in the HLT. Information from the muon chambers, calorimeters (m.i.p energy deposits) and tracking system are used to fit the muon track throughout the full detector.

The data from these streams were used to produce calibration and alignment constants used by the offline data reprocessing.

In order to monitor the performances of the HLT reconstruction and selection, an additional stream was implemented. The so called "*HLTDebug*" stream stores, for every accepted event, all the relevant information reconstructed by the HLT, in particular the physics candidates which the HLT selection is based upon. The stream was fed by the same paths forming the physics stream plus additional pre-scaled paths selecting the other types of L1T triggers (calibration and random). This was possible due to the reduced trigger rate and data volume. Clearly during LHC operation when the bandwidth to disk becomes critical, only a subset of the trigger paths can be duplicated for debugging purposes.

During the commissioning of the detector, data were also collected in non standard conditions, for example while the magnetic field was ramping up. Such conditions may lead to pathological events which the reconstruction software may not handle properly. In the case of a failure during the execution of the HLT, the corresponding Event Processors were recovered and the raw detector data sent to a dedicated *Error* stream (cf section 3.6).

The high rate random triggered events were also used to verify the bandwidth to disk capacity of the storage system. To perform this test, the prescale factor applied initially on the random triggers was reduced until saturation of the HLT output was achieved. A system reduced to one forth of the final designed setup was used, with 4 SM's connected to 2 SATABeast$^{TM}$ disk arrays. The results are summarized by the plot of Figure 9, showing the data rate to disk as a function of time for the two cosmic rays run when the stress test was performed.

The overall bandwidth to disk was above 600 MB/s. The fluctuations are due to the data transfer to the CERN Tier-0 which consumes part of the bandwidth available for writing. The shaded band accounts for the fact that the utilized bandwidth for reading was only known approximately. The plot displays separately the contributions of the data streams implemented in the HLT configuration: as expected, the *HLTDebug* stream accounts for the largest part. No calibration and alignement streams were implemented during those runs, in particular the calibration triggers were not enabled during those runs.

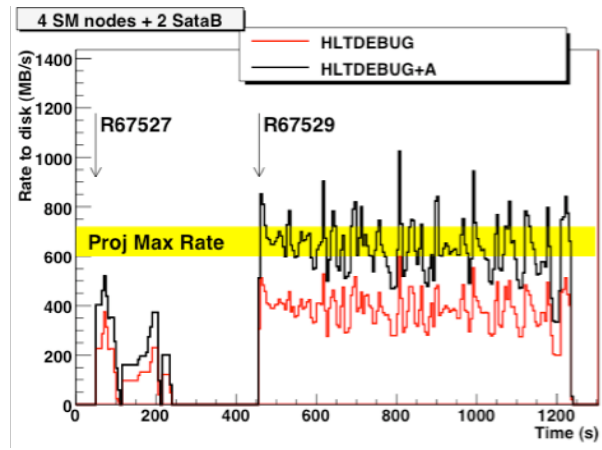

**Figure 9.** Rate of data to disk as a function of time for two cosmic rays runs where the high rate randomly triggered events were only partially filtered out by the HLT, thus saturating the bandwidth of the storage system. The contributions from the two data streams implemented during those runs are shown separately.

*6.3 Data Quality Monitoring*

The online DQM was extensively utilized during the commissioning runs as a fundamental tool for the identification and the debugging of issues regarding both the various detector components and the trigger system.

The DQM applications run at both the stages described in section 3.7. Having access to all the L1-accepted events is desirable for data integrity checks and to monitor the performance of the L1 and HLT operations, the corresponding modules were therefore embedded into the HLT configuration. The plot in figure 10 is an example of a HLT DQM histogram showing the rate (color code in the z-axis) at which a given module instance (y-axis) along a given path (x-axis) is executed.

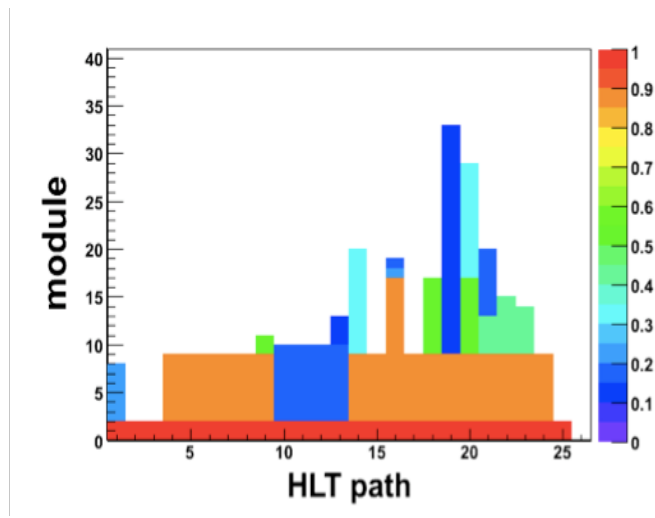

**Figure 10.** Example of DQM histogram for the monitoring of the HLT performances. The HLT paths are listed on the horizontal axis, whereas the vertical axis represents the depth along the reconstruction sequence. The color code expresses the rate at which a given reconstruction module is executed. This result is for a cosmic muon run with dedicated HLT menu.

Many of the online DQM applications, dedicated to the monitoring of the detector data, were fed by the SM proxy server which provided events from the *HLTDebug* stream. Events from the latter already contained the reconstructed objects on which the quality of the data could be determined.

For each run, the results of the DQM applications were compared and then used to classify the overall quality of the run.

## 7. HLT operations with the first LHC beams

In summer 2008, the accelerator team started injecting 450 GeV proton beams into the main LHC ring. CMS detected the first beam induced activity at the beginning of September 2008, when a beam was steered onto collimators placed upstream of the CMS detector. On September 10, 2008, and over the course of the following nine days, two counter rotating proton beams have circulated (one at the time) along the entire LHC ring, passing several times through CMS. Due to the unfortunate incident occurring in the sector 3-4 of the LHC, the collider operations have been suspended on September 19, 2008, before achieving the first proton-proton collisions. "LHC beam operations are scheduled to resume in September 2009.

The detection of the first LHC beam induced events [20] has been very useful for the commissioning of the whole experiment, in particular to study the timing of the trigger system and of the signal capture by the various sub-detectors. The trigger and DAQ system has been configured to record single beam events. Two types of L1 triggers have been employed:

- Beam halo triggers, used to trigger on the detector activity induced by the beam, such as beam halo muons passing through the muon end-cap (ME) chambers and through the forward hadron calorimeters (HF). The muon halo trigger rates were between few Hz till up to O(100) Hz at the de-bunching of the beam (see Figure 11);
- Beam capture triggers, based on the signals from beam detectors, as the Beam Pickup (BPTX), placed along the beam pipe, at various distances from the CMS detector. These devices were providing a trigger signal at every beam passage with a trigger rate as high as 11 kHz in the case of a (single) continuously circulating beams.

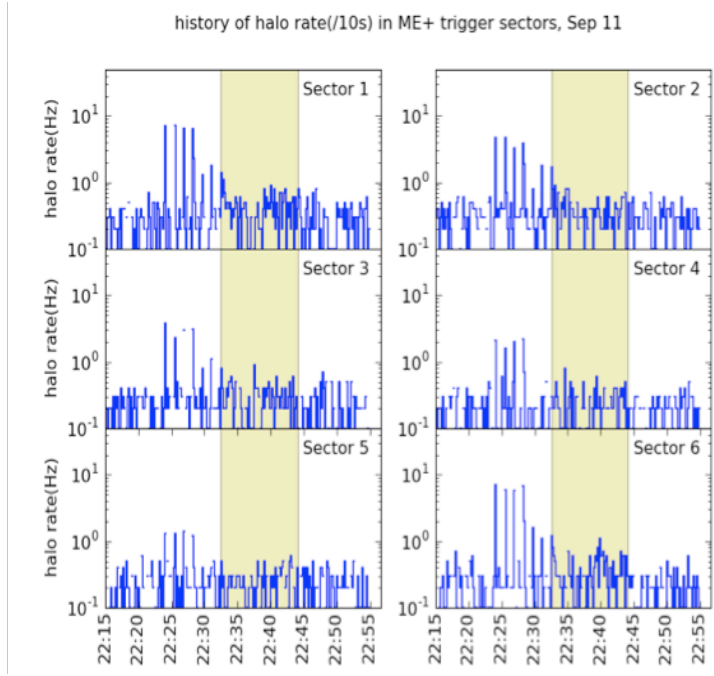

**Figure 11.** Muon halo rates recorded in the muon end-cap (ME) trigger sectors, at the passage of the beam through CMS. The yellow band shows the beam capture signal recorded downstream of the CMS detector.

The HLT system was configured to apply no rejection, but simply "tag-and-pass" the events accordingly to the L1 information. In the case of continuously circulating beam, the beam capture trigger was pre-scaled directly at L1, thus reducing the HLT rate to a level sustainable by the storage system. Each of the paths implemented in the HLT menu was seeded by a specific L1 bit. Thus, the HLT unpacking of the raw data was performed on each L1 accepted event.

Beam capture L1 triggers were seeding also two dedicated HLT paths, applying special filters to detect beam-induced signals in the central detectors. These filters were based on very simple quantities, such as the size of the data payload of the pixel vertex detector (actually kept off during the first runs with the circulating beam) and the energy deposition measured by the calorimeters. These filters required no reconstruction to be performed, except for the calorimeter local reconstruction.

The data taking was flawless and efficient during the entire period of beam runs. The BPTX signal was synchronized with the trigger system almost immediately (at the third beam shot of beam 1 to CMS). From that moment on, the events were continuously logged on to disk. Beam events selected by the HLT were immediately made available to the online DQM and the Event Display. The histograms produced in the Filter Farm by the DQM applications have proven very useful to monitor the trigger performances and their evolution with the time. Examples of online event displays are reported in Figure 12 and 13.

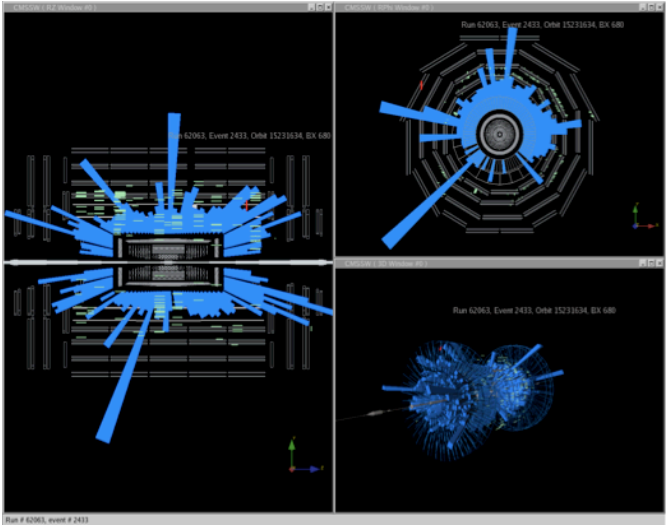
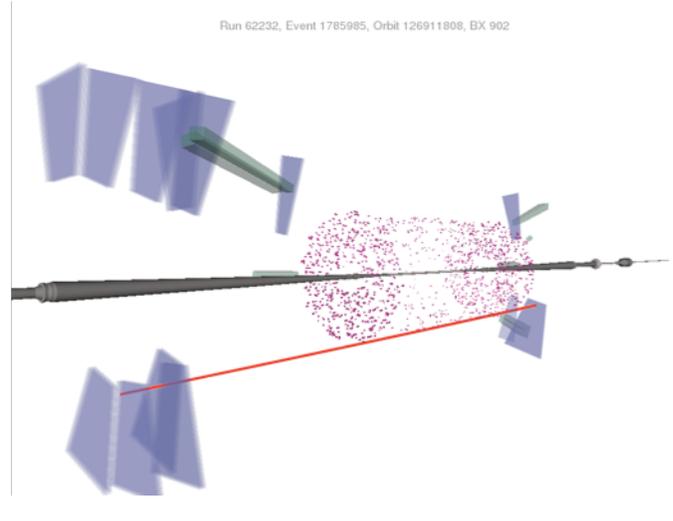

**Figure 12.** Beam splash event: muons originated from ($2 \times 10^9$) protons hitting the collimator blocks situated about 150 m upstream of CMS, were detected in the calorimeters (calorimeter deposits shown in blue) and muon chambers (muon chamber hits shown in green).

**Figure 13.** Sections of the CMS detector displaying a beam halo muon.

The event display in Figure 12 shows the effect of the beam smashing on the collimators located upstream of the CMS detector: hundreds of thousands of particles passing through the detector deposit a large amount of total energy (equivalent to ~1000 TeV) in the hadron calorimeter. The event display in Figure 13 shows a beam halo muon, crossing CMS from one end-cap side to the other, and reconstructed in the muon chambers of both end-cap disks.

**8. Summary and outlook**

As of summer 2008, the CMS detector is built and undergoing its final commissioning phase to get ready for LHC collisions. The two level trigger system is a key design feature of the experiment, with the second level, the High Level Trigger, due to reduce the Level 1 output rate from 100 kHz to 0(100) Hz. The features of the CMS High Level Trigger have been described. The performances of the system during the cosmic rays data taking well matched the expectations. HLT menus of increasing complexity have been deployed in the online computer farm and proved to run stably for several hours. Input rate as high as 60 kHz can be managed by the DAQ system, whereas the bandwidth to disk reaches 600 MB/sec with only one fourth of the storage system foreseen for the final setup. CMS collected very efficiently good quality data during the first LHC beam operations with the HLT having the task of selecting and tagging events for prompt monitoring and analysis.

**Acknowledgments**

The results reported in this paper could have not been achieved without the fundamental contributions from our colleagues of the CMS collaboration. We would like to acknowledge all of them for their efforts and for turning an ambitious project into an extraordinary scientific device.